# Turbulence Driving by Interstellar Pickup Ions in the Outer Solar Wind

Philip A. Isenberg, Bernard J. Vasquez, and Charles W. Smith
Space Science Center and Department of Physics and Astronomy
University of New Hampshire, Durham, NH 03824, USA

**Abstract**. We revisit the question of how the unstable scattering of interstellar pickup ions (PUIs) may drive turbulence in the outer solar wind, and why the energy released into fluctuations by this scattering appears to be significantly less than the standard bispherical prediction. We suggest that energization of the newly picked-up ions by the ambient turbulence during the scattering process can result in a more spherical distribution of PUIs, and reduce the generated fluctuation energy to a level consistent with the observations of turbulent intensities and core solar wind heating. This scenario implies the operation of a self-regulation mechanism that maintains the observed conditions of turbulence and heating in the PUI-dominated solar wind.

## 1. Introduction

The neutral component of the partially ionized local interstellar medium (LISM) penetrates the boundaries of the heliopause and termination shock and flows slowly through the supersonic solar wind (Axford 1972; Holzer 1972; Isenberg 1986). These neutral particles, primarily hydrogen atoms, fill most of the heliosphere, being excluded only from a region near the Sun due to increased ionization from solar photons and charge-exchange interactions with the solar wind protons. As the solar wind expands away from the Sun, the interactions between the streaming plasma and the relatively stationary LISM atoms become energetically important, resulting in a deceleration and heating of the solar wind, as well as significant modifications of the termination shock.

The detailed interaction begins when an LISM atom is ionized, creating a new ion that is temporarily nearly at rest with respect to the supersonic solar wind. In the reference frame of the solar wind plasma, this new ion is streaming toward the Sun at the solar wind speed, $V_{sw}$. Suddenly subject to the electromagnetic fields of the flowing plasma, the new ion is "picked up" as its motion across the local magnetic field, **B**, is forced into a gyration with a speed perpendicular to the field equal to $V_{sw} \sin \Psi$, where $\Psi$ is the angle between **B** and **V**$_{\mathbf{sw}}$. The distribution of these ring-beam particles is highly unstable, and will generate parallel-propagating ion-cyclotron waves (ICWs), while scattering to a nearly isotropic distribution of hot pickup ions (PUIs) around a cooler proton core of solar origin.

Early consideration of this scattering process predicted a substantial ICW enhancement at frequencies above the spacecraft frame proton gyrofrequency due to the PUI isotropization (Wu & Davidson 1972; Lee & Ip 1987). However, these waves were only observed sporadically (Murphy et al. 1995), despite the expected continual production of new PUIs. Eventually, it became clear that the predicted waves were being dispersed and masked by the background turbulence in the outer solar wind (Cannon et al. 2014a; Cannon et al. 2014b; Joyce et al. 2010;

Aggarwal et al. 2016; Argall et al. 2015; Argall et al. 2017; Fisher et al. 2016; Marchuk et al. 2021). The distinctive wave enhancements would appear during quiet periods in the solar wind when the level of ambient turbulence was low, but were not detectable during times of stronger turbulence (Hollick et al. 2018a, b; Pine et al. 2020a).

These ICWs are also presumed to drive the turbulent motions in the expanding solar wind, supplementing and eventually replacing the standard energy inputs from shocks and shears that diminish far from the Sun. This turbulent driving by PUIs was first proposed by Zank, Matthaeus, and co-workers (Matthaeus et al. 1999; Williams et al. 1995; Zank et al. 1996). They presented a model for turbulent evolution from the Sun to the termination shock, based on the "engineering model" concepts of von Kármán & Howarth (von Kármán & Howarth 1938; Kolmogorov 1941a, b). Within several AU of the Sun, this model turbulence was driven by a phenomenological shear source, and further away incorporated the energy released by PUI isotropization. The dissipation of this turbulence was taken to heat the core solar wind protons, which measurements from *Voyager 2* showed were not cooling adiabatically (Gazis & Lazarus 1982; Gazis 1984).

The early models predicted that the core solar wind temperatures would increase dramatically beyond ~ 40 AU where the PUI driving finally dominated the adiabatic cooling. This prediction arose from the assumption that new PUIs would scatter into a "bispherical" distribution as a result of their instability (Galeev & Sagdeev 1988; Johnstone et al. 1991; Williams & Zank 1994). This distribution (described in the next section) corresponds to a reduction of the particle energy per mass by a factor ~ $V_A/V_{sw}$ from the initial $V_{sw}^2/2$ of the newly ionized ring-beam in the solar wind frame, where $V_A$ is the local Alfvén speed. In these models, the energy lost by each newly ionized pickup proton was added to the turbulent fluctuations as the solar wind continued outward, and the dissipation of this turbulence produced a temperature increase with increasing radial distance from the Sun.

However, as *Voyager 2* traveled further outward, the measured proton temperature showed only a modest increase, at a fraction of the predicted magnitude (Smith et al. 2001). We proposed an explanation for this discrepancy, suggesting that half of the unstable ICWs could be transformed by the ambient turbulence into parallel-propagating fast-mode waves as they were being generated by the instability (Isenberg et al. 2003; Isenberg 2005). Cyclotron-resonant scattering by the combined ICWs and fast-mode waves led to a distribution of new PUIs which was much more isotropic than the bispherical distribution, so less energy was supplied to the turbulence. Models of the distant solar wind under this hypothesis could reproduce the *Voyager 2* observations reasonably well, even under time-dependent solar wind conditions (Isenberg et al. 2010; Smith et al. 2006).

Since that work, many further models of increasing rigor and complexity have been constructed (Adhikari et al. 2014; Adhikari et al. 2017; Adhikari et al. 2020; Zank et al. 2018; Oughton et al. 2011; Usmanov & Goldstein 2006; Usmanov et al. 2011; Usmanov et al. 2012, 2014, 2016; Pine et al. 2020d, c, b; Pine et al. 2020a; Ng et al. 2010; Wiengarten et al. 2016), exploring the evolution and behavior of various turbulent quantities as the solar wind advects into the outer heliosphere. Of course, the more elaborate models contain more free parameters, so a decent agreement with the *Voyager 2* observations has been maintained. The newer measurements from the Solar Wind around Pluto (SWAP) instrument on *New Horizons*, (McComas et al. 2008) which includes information on the PUI component, are also generally consistent with the behavior of these models (McComas et al. 2021; Elliott et al. 2019; McComas et al. 2017; Zank et al. 2018).

In all this more recent work, the specific character of the PUI driving has not been addressed. In fact, most of the models have retreated from the detailed kinetic model of Isenberg (2005), and simply taken the fractional decrease of the driving from the bispherical value to be a constant, usually designated $f_D$. In this context, values of $f_D$ between 0.15 - 0.25 have led to plausible agreement with the observations.

In this paper, we revisit the question of PUI driven turbulence and take a new look at how the isotropization process could release less energy than the bispherical expectation. We note that, in the years since our suggestion of fast-mode PUI scattering, there has been no evidence that fast-mode waves play a major role in the establishment of strong plasma turbulence (Yang et al. 2018; Brodiano et al. 2021). That original hypothesis may have seemed plausible in the rigid context of resonant scattering of unstable PUIs by strictly parallel-propagating waves, but newer conceptions of solar wind turbulence have prompted another look at this question.

Here, we will explore the possibility that the ambient turbulence in the outer solar wind reduces the energy lost from isotropizing interstellar PUIs, and self-consistently modulates the subsequent driving of the cascade, dissipation and heating. The simultaneous and comparable kinetic influence of quasilinear pitch-angle scattering and turbulent dissipation is not a familiar concept in the magnetosphere or inner heliosphere, where the plasmas are more active and the unstable ion gradients are often well established. In the outer solar wind, however, the production of newly ionized PUIs is very slow and their unstable pitch-angle gradient is only weakly maintained. Thus, the operation of the ring-beam instability, which would scatter the new PUIs to a bispherical shell, can be modified by other kinetic processes in the plasma. In particular, the ambient turbulent fluctuations themselves could influence the PUI isotropization process as they dissipate and heat the rest of the plasma.

This point is consistent with the prevailing explanation for the sporadic nature of the observed PUI excited waves in the outer solar wind (Cannon et al. 2014b; Hollick et al. 2018b; Pine et al. 2020a). Those observations have shown that the turbulent cascade rate is roughly comparable to the PUI production rate, and that PUI excited wave enhancements are only evident when the cascade rate is too small to disperse them. Here, we extend this picture to suggest that the turbulence heats the PUIs as they scatter, interfering with the formation of the bispherical shell.

In the next section, we describe the wave-particle interactions and the solar wind conditions that are relevant to the isotropization of PUIs. In Section 3, we present an idealized demonstration of how combined PUI scattering and turbulent diffusion can modify the energy available for turbulence driving in the outer solar wind. In Section 4, we outline how these combined processes can balance each other, creating a self-regulating system to maintain the levels of turbulence and heating consistent with observations. Section 5 contains a summary and our conclusions.

## 2. PUI Isotropization in the Outer Solar Wind

The ring-beam instability of newly picked up ions has been extensively studied in many different contexts (Lee & Ip 1987; Galeev & Sagdeev 1988; Min et al. 2017; Min & Liu 2018; Cowee et al. 2008; Huddleston et al. 1992; Huddleston et al. 1998; Szegö et al. 2000). The resulting waves have been observed at comets, in planetary magnetospheres, and in the solar wind. Further studies in the context of the IBEX ribbon in the outer heliosheath have been focused on the possibility that this wave generation might be somehow prevented in that very quiet medium, in order to allow the accumulation of ribbon ions (Florinski et al. 2010; Florinski et al. 2016; Liu et al. 2012; Summerlin et al. 2014). At this time, an inhibition sufficient to justify that hypothesis has yet to be demonstrated.

We do not have direct observations of this instability in the distant supersonic solar wind, in that we cannot detect PUI gradients and resonant ICWs simultaneously. However, the persistent operation of this instability to isotropize interstellar PUIs has been indirectly verified by the observation of wave enhancements in the measured spectra (Cannon et al. 2014a; Cannon et al. 2014b; Hollick et al. 2018a, b; Pine et al. 2020d, c), by the analysis of isotropic PUI distributions at *New Horizons* (Elliott et al. 2019; McComas et al. 2017; McComas et al. 2021), and by the increasing temperature of core solar wind protons plausibly heated by the turbulence as it is continuously driven by the instability (Adhikari et al. 2017; Isenberg et al. 2010; Smith et al. 2006; Usmanov et al. 2014).

The ring-beam instability is a form of the well-known quasilinear (QL) anisotropy instability (Gary 1993) which occurs when ions exhibit a pitch-angle gradient perpendicular to the large-scale magnetic field. The gradient represents a form of free energy, which generates ICWs propagating primarily along the field. In the generation process, these waves self-consistently scatter the cyclotron resonant ions to reduce the pitch-angle gradient. The final bispherical shell configuration results from the details of the QL resonant cyclotron interaction with these waves.

The cyclotron resonance condition for an ion in a transverse wave is written

$$\omega - k_{\parallel} v_{\parallel} = \Omega_i , \qquad (1)$$

which arises from equating the ion cyclotron frequency, $\Omega_i$, with the Doppler-shifted wave frequency encountered by the ion streaming at $v_{\parallel}$. With the appropriate polarization, the wave fields rotate about the large-scale magnetic field in phase with the ion gyration, and energy exchange between the wave and the ion is efficient. A resonant ion viewed in the reference frame propagating with the parallel phase speed of the wave, $\omega/k_{\parallel}$, will see a zero wave electric field, since $dB/dt = 0$ in that frame. Thus, the ion will interact with the wave so as to conserve its energy in that wave frame. Back in the plasma frame, this means that resonant ions will scatter through their phase space along a particular curved path, determined by the initial ring-beam and the wave dispersion relation, $\omega(k)$. In the distant solar wind, the Parker spiral field, on average, is azimuthally-directed ($\Psi = \pi/2$) so the initial PUIs will appear in the ring at $v_{\perp} = V_{SW}$ and $v_{\parallel} = 0$.

If the resonant waves can be taken dispersionless, $\omega/k_{\parallel} = \pm V_A$, this energy-conserving path follows symmetric spherical sections in $(v_{\parallel}, v_{\perp})$ space with centers at $(\pm V_A, 0)$, and the final closed ion shell has the standard bispherical shape, as shown by the blue curve in Figure 1. Including the effects of ICW dispersion will modify this shape, typically making it somewhat more energetic since the resonant waves will propagate slower and less energy is taken from the scattering ions. (As a limiting case, consider the diagram in Figure 1 for a wave phase speed equal to zero. In this case, the center of curvature of each path would be at the origin, and the scattered shell shape in the plasma frame would be a sphere. In this limit, the PUIs would retain all their initial energy and yield none to the waves.)

An analytical solution including dispersion can be obtained by taking the dispersion relation for parallel ICWs in a cold proton-electron plasma, $\omega = k_\parallel V_A \sqrt{1 - \omega/\Omega_p}$, resulting in a scattered shell shape given by the parametric expressions

$$v_\perp^2 = V_{sw}^2 - V_A^2 \left[ \frac{1}{y^2} + \ln y - \sinh^{-1}\frac{y}{2} \right] \tag{2}$$

$$v_\parallel = V_A \left[ \frac{1}{y} + \frac{\sqrt{y^2 + 4} - y}{2} \right]$$

where $y = k_\parallel V_A/\Omega_p$. This more realistic shape, termed a "dispersive bispherical" by Isenberg & Lee (1996), corresponds to less generated wave energy than the standard bispherical, but the reduction is not large enough to resolve the outer solar wind heating problem on its own.

As stated above, turbulent heating from waves generated by PUI isotropization is plausibly responsible for the observed core solar wind temperatures in the outer solar wind, but only if this wave generation is somehow reduced by a factor of 5 or more from the expected QL values. These QL calculations are made in isolation, assuming an otherwise undisturbed, homogeneous background plasma. In the outer solar wind, though, this isotropization must proceed in the presence of a well-developed turbulence, which can act simultaneously to energize the scattering PUIs. We suggest here that this energization is responsible for the reduced wave generation required by the models. This effect is shown schematically by the red dashed curve in Figure 1.

In the outer solar wind, the PUI anisotropy as such is maintained only by the production of new PUIs appearing in the ring position at $v_\perp = V_{sw}$. The proton production rate is given by the local density of interstellar neutral hydrogen multiplied by their ionization rate

$$\frac{dN}{dt} = N_o \exp(-L/r) \, v_o \left( \frac{r_o}{r} \right)^2 \text{ cm}^{-3} \text{ s}^{-1}, \tag{3}$$

where $N_o = 0.1$ cm$^{-3}$ is the inflowing hydrogen density at the solar wind termination shock, $L = 5.6$ AU is the size of the hydrogen ionization cavity in the upwind direction, and $v_o = 7.5 \times 10^{-7}$ s$^{-1}$ is the hydrogen ionization rate at $r_o = 1$ AU taken to fall off as $r^{-2}$. At $r = 40$ AU, we estimate this rate to be $\sim 4 \times 10^{-11}$ particles cm$^{-3}$ s$^{-1}$, so about one new proton/m$^3$ every 7 hours. In contrast, the proton gyroperiod at this distance will be about 1 minute or less.

The associated rate per mass of energy input to the fluctuations is then

$$\frac{dE_w}{dt} = \frac{1}{2} f_D V_A V_{sw} \frac{dN}{dt}, \tag{4}$$

where we follow Smith et al. (2001) in parameterizing the effective fraction of bispherical wave generation by the factor $f_D$. Taking $V_{sw} = 400$ km/s and $V_A = 50$ km/s at $r = 40$ AU, the energy input rate is $dE_w/dt = 4 \times 10^{-7} f_D$ (km/s)$^2$ cm$^{-3}$ s$^{-1}$.

This energy input rate should be compared with the rate of energy cascade through the inertial range of the turbulent spectrum. This rate is estimated by Vasquez et al. (2007) from the Kolmogorov theory to be

$$\varepsilon_K = \frac{f_{sc}^{5/2}\left[E(f_{sc})\right]^{3/2}(21.8)^3}{V_{sw}n_p^{3/2}} \text{km}^2\text{s}^{-3} \quad , \tag{5}$$

where $E(f_{sc})$ is the magnetic field power spectral density in units of nT$^2$/Hz, which is assumed to vary with the measured spacecraft-frame frequency, $f_{sc}$, as $f_{sc}^{-5/3}$, $n_p$ is the solar wind density in units of cm$^{-3}$, and the factors of $n_p$ and 21.8 convert the magnetic field to Alfvén units. At 40 AU, we estimate $E$ at $f_{sc}$ = 3 mHz from Figure 4 of Pine et al. (2020a) to be $10^{-2}$ nT/Hz. We multiply the expression in (5) by the solar wind density $n_p = 3.125 \times 10^{-3}$ cm$^{-3}$ (corresponding to 5 cm$^{-3}$ at 1 AU) to obtain a cascade rate per volume of $2.28 \times 10^{-7}$ (km/s)$^2$ cm$^{-3}$ s$^{-1}$. In steady state, the turbulent dissipation rate is equal to the cascade rate, so the solar wind protons including the PUIs will be heated by this rate - comparable to the turbulent driving rate (4) due to PUIs.

The order-of-magnitude equality of these two rates is one of the foundational factors of the prevailing scenario for turbulence driving and solar wind heating in the outer heliosphere. Observational studies (e.g. Cannon et al. 2014b; Hollick et al. 2018b; Pine et al. 2020d; Pine et al. 2020a) have found that the PUI-generated waves due to isotropization, as predicted by Lee & Ip (1987), are observable when the level of background turbulence is low, and are not detected during times of higher turbulence. Since interstellar PUIs are ionized and picked up continuously, independent of the solar wind turbulence level, it is understood that the isotropization continues to take place, and that the turbulence continues to be driven by this process.

Previous models have all assumed that the isotropization process is rapid and not affected by the turbulence. However, the proton gradient assumed to generate the anisotropy instability will not be firmly established when new ring distribution ions only appear every few hours. Although these new ions apparently do scatter toward isotropy, since they do still drive the turbulence, this scattering cannot be rapid and is not likely to be independent of other ongoing processes.

The interplay between anisotropy instabilities and background turbulence has been investigated in various simulation studies in the magnetosphere and inner heliosphere (Hellinger et al. 2015; Hellinger & Trávnícek 2015; Markovskii et al. 2019, 2020; Markovskii & Vasquez 2022). The typical conclusion is that, while the growth rates and saturation levels of the instability may be affected, the unstable wave-particle evolution is not strongly suppressed by the turbulence.

Additionally, several hybrid simulations (Hellinger & Trávnícek 2016; Liu et al. 2012) have demonstrated the continued operation of this instability under conditions of slow PUI production rate. Unfortunately, neither study was concerned with the shape of the closed shells that resulted from the scattering or with the energy lost by the scattered particles. Furthermore, both these studies were spatially one-dimensional, so all fluctuations were forced to propagate along the magnetic field, and true turbulence could not develop.

Previous models of turbulence in the outer solar wind have also assumed that PUIs would not be heated by the turbulent dissipation. It is generally argued (e.g. Zank et al. 2018) that a resonant heating of suprathermal ions requires large-wavelength fluctuations, while dissipation-range turbulence is characterized by very short wavelengths. However, to the extent that the dominant turbulent fluctuations can be represented as highly oblique kinetic Alfvén waves in critical balance, the short perpendicular scales in the dissipation range can easily coexist with larger parallel scales (Goldreich & Sridhar 1995, 1997; Isenberg & Vasquez 2019; TenBarge & Howes 2012; Oughton et al. 2015). In any case, we also point out that the PUIs observed at *New Horizons* in the outer solar wind appear to be heated in situ (Elliott et al. 2019; McComas et al. 2021; Zank et al. 2018) in a manner that has not been fully explained.

In the next section, we present an idealized demonstration that is meant to illustrate the effects of slow QL scattering by the instability combined with ambient heating by dissipation of the background turbulence.

### 3. An Idealized Demonstration

A rigorous demonstration of these combined processes would require a full kinetic simulation of very slow PUI production in an otherwise turbulent plasma, which is beyond the scope of this paper. We present here an idealized demonstration of how this interaction might work, using particle diffusion operators to represent the processes of scattering and heating.

This idealization has a number of limitations. Since diffusion only produces a directed result - such as scattering toward isotropy or net ion heating - in response to a directed density gradient, we need to impose a specific background particle distribution. Additionally, we will specify diffusion coefficients for the two processes, but these effects will then be independent and will not be able to affect each other.

We will represent the QL scattering using the "dispersive bispherical" description of Eq. (2), and model this process on the very slow time scale by taking the magnitude of this diffusion coefficient to be very small. In the absence of a clearly accepted form for the dissipative turbulent heating, we will choose the diffusion coefficient for heating to be a simple isotropic function of particle speed. Then, to assess the evolution of newly ionized PUIs, we will insert a "slug" of additional protons in a ring at the solar wind speed and compare the energization of the total distribution to that without the slug.

With this procedure, the pickup of newly ionized protons is not continuous, but rather the new particles are dumped into the system at once. We do not compute the QL anisotropy instability or the self-consistent particle response. Rather, we define the QL diffusion to follow the dispersive bispherical path through phase space, taking this diffusion to be slow.

Our demonstration will also neglect the effects of adiabatic deceleration in the expanding solar wind, which proceeds on an even longer time scale than the particle effects of interest. This neglect means that the distinctive solar wind speed cutoff in the PUI distributions, to be imposed on the initial distribution, will not be maintained in this model. Consequently, the kinetic behavior of the model PUI distribution is not expected to be physically realistic after the initial pitch-angle scattering time period.

Taking a grid in spherical coordinates ($\mu$, $v$), where $\mu$ is the cosine of the particle pitch angle, we solve the two-dimensional time-dependent diffusion equation

$$\frac{\partial f}{\partial t} = \frac{\partial}{\partial \mu}\left(D_{\mu\mu}\frac{\partial f}{\partial \mu} + D_{\mu v}\frac{\partial f}{\partial v}\right) + \frac{1}{v^2}\frac{\partial}{\partial v}\left[v^2\left(D_{\mu v}\frac{\partial f}{\partial \mu} + D_{vv}\frac{\partial f}{\partial v}\right)\right] \qquad (6)$$

in a system symmetric about $\mu = 0$.

The QL diffusion coefficients for the cyclotron resonant scattering due to parallel-propagating ICWs are (Isenberg 2005; Lee 1971; Schlickeiser 1989)

$$\left\{\begin{array}{c} D_{\mu\mu} \\ D_{\mu\upsilon} \\ D_{\upsilon\upsilon} \end{array}\right\} = A_{IC} \frac{(1-\mu^2)I(k_{res})}{|\mu\upsilon - V_{gr}|} \left\{\begin{array}{c} \left(1 - \frac{\mu V_{ph}}{\upsilon}\right)^2 \\ V_{ph}\left(1 - \frac{\mu V_{ph}}{\upsilon}\right) \\ V_{ph}^2 \end{array}\right\} \quad (7)$$

where the spectral intensity of the ICWs at the resonant wavenumber is $A_{IC} I(k_{res})$, and their phase and group speeds are $V_{ph}$ and $V_{gr}$, respectively. For diffusion coefficients specific to PUI scattering in an azimuthal magnetic field, we take the shape of the ICW spectrum from the dispersive bispherical formalism (Isenberg & Lee 1996) for perpendicular pickup. The wave-particle interaction is assumed symmetric about $\mu = 0$, so the spectrum is taken symmetric about $k = 0$.

The turbulent heating here is modeled by an isotropic power-law diffusion in $\upsilon$, with a Gaussian decrease to zero in the diffusion coefficient for small $\upsilon$,

$$D_{\upsilon\upsilon} = A_{turb} \left\{\begin{array}{ll} (\upsilon/V_A)^{-5/3} & \upsilon/V_A \geq 1 \\ \exp\left[-\frac{(\upsilon-V_A)^2}{(V_A/5)^2}\right] & \upsilon/V_A \leq 1 \end{array}\right. \quad (8)$$

The total diffusion coefficients in (6) will be the sums of the expressions in (7) and (8), where $A_{IC} I(k_o = \Omega/V_A)$ and $A_{turb}$ are constants to be set for various runs.

We perform these computations on a grid with 2000 x 4000 points between $0 \leq \mu \leq 1$ and $0 \leq \upsilon/V_A \leq 10$. As appropriate for the outer solar wind, where the magnetic field is essentially azimuthal, we assume that new PUIs appear near $\mu = 0$, $\upsilon = V_{sw}$, and that the particle distribution is symmetric about $\mu = 0$. We take $V_{sw}/V_A = 6$ rather than a more typical value ~ 10, in order to reduce the necessary computational volume in phase space.

The background proton distribution is composed of a thermal core with the ratio of kinetic to magnetic pressure $\beta = 0.2$, a halo of previously picked up PUIs in a Vasyliunas & Siscoe (1976) configuration, and a sharp cutoff above $\upsilon = V_{sw}$:

$$f(\upsilon) = \left\{\begin{array}{ll} \exp\left[-\frac{5\upsilon^2}{V_A^2}\right] + b\left(\frac{\upsilon}{V_{sw}}\right)^{-3/2} \exp\left[-0.1\left(\frac{\upsilon}{V_{sw}}\right)^{-3/2}\right] & \upsilon \leq V_{sw} \\ b\left(\frac{\upsilon}{V_{sw}}\right)^{-30} \exp(-0.1) & \upsilon \geq V_{sw} \end{array}\right. \quad (9)$$

where $b$ is chosen to give a PUI halo density that is 10% of the total. This distribution is shown in Figure 2.

With the initial condition given by (9), we solve (6) on the grid using the biconjugate gradient stabilized method. We take reflecting boundary conditions at $\mu = 0$ and 1, and place an absorbing boundary at $\upsilon = 10\ V_A$. Given the presence of this outer boundary in phase space, along with the neglect of adiabatic deceleration, we expect this idealized model to only yield physically reasonable distributions for a period near the beginning of the computation. The results for an initial time period, which are shorter for larger values of $A_{turb}$, will be qualitatively illustrative of the combined scattering and diffusive heating of newly ionized PUIs in the outer solar wind.

We first consider the case of $A_{turb} = 0$, which should reproduce the conditions treated in the original dispersive bispherical calculation (Isenberg & Lee 1996). In keeping with the slow time scale for pickup in the outer solar wind, we take $A_{IC} I(k_o) = 10^{-5}$ in computational units.[1] Starting with the background distribution (9), we follow the total kinetic energy of the protons as a function of time, $E_b(t)$, which is shown by the blue curve in Figure 3.

We then repeat this computation starting with an additional slug of particles, representing newly ionized PUIs. These additional PUIs are confined to the pitch angle range $0 \leq \mu \leq 0.03$ and the speed range $5.85 \leq v/V_A \leq 6.05$, which straddles the model solar wind speed at $V_{sw} = 6 V_A$. In this region, the we take the value of $f$ to be a constant equal to the value from (9) at $v = 5.85 V_A$. With this shelf of extra particles in the distribution, the total proton density is increased by a factor of $10^{-4}$. The total proton energy in this case, $E_s(t)$, is shown by the red curve in Figure 3.

To quantify the energetics of QL scattering of the newly ionized PUIs, we compare the evolution of the total energy of the proton distributions in these two computations. Ideally, one might prefer to track the actual scattered particles, but in this diffusive system the discernable density enhancement in the slug quickly disappears and we are unable to follow the behavior of that small portion of the distribution.

Figure 4 shows the quantity $Q \equiv \Delta E(t)/\Delta E(0)$, where $\Delta E = E_s - E_b$ for this case. In this context, $(1 - Q)$ represents the normalized net energy lost by the new PUIs as they scatter toward isotropy in the undisturbed QL interaction. The normalized energy change levels off at $Q = 0.893$, which is the same value predicted by the dispersive bispherical calculation of Isenberg & Lee (1996) for $V_{sw}/V_A = 6$. In the turbulent driving scenario for the outer solar wind (Isenberg et al. 2003; Isenberg 2005; Smith et al. 2001; Smith et al. 2006; Zank et al. 1996; Pine et al. 2020a), this proton energy loss appears in ICWs that proceed to drive the turbulence, but the observations at *Pioneer 11* and *Voyager* 2 show that this amount of driving is too large.

When turbulent heating is added to the system, the combined 2-D diffusion proceeds somewhat faster than the effectively 1-D pitch-angle scattering that occurs when $A_{turb} = 0$. The total proton energy rises approximately linearly with time for all cases. Keeping $A_{IC} I(k_o) = 10^{-5}$, we choose five values of $A_{turb} = 0.0001, 0.0005, 0.001, 0.002,$ and $0.003$. In Figure 5, we show the results of the normalized energy difference, $Q(t)$, between the computations made with and without the slug of new PUIs for these cases. The computed energy difference reaches a minimum depending on the value of $A_{turb}$, before increasing due to further diffusive heating. We interpret these minimum values to indicate the normalized energy of the slug once it has completely scattered to a closed shell in phase space, corresponding to the end state of the QL anisotropy instability. The curves in Figure 5 show that the cases with larger values of $A_{turb}$ lose less energy in the process of scattering. The minimum values for each case are given in Table 1.

---

[1] This illustrative computation only addresses the relative effects of isotropization and turbulent heating. As such, it is sufficient to work in dimensionless computational, or grid, units. All velocities are normalized to $V_A$, and total proton kinetic energies per proton mass are then defined as $\frac{1}{2} f \times (v/V_A)^4$, summed over all grid points. The time parameter then scales as $V_A (\Omega A_{IC} I(k_o))^{-1}$.

These computations suggest that the resolution to the discrepancy between the predicted energy loss of new PUIs and the observed turbulent heating of the core solar wind can be found in the influence of the background turbulence on the weak QL scattering of these PUIs. In this example, the new PUIs can be scattered to a stable closed shell while giving up a fraction of their energy, which can range from 11% down to 1%, depending on the strength of the background turbulence. Defining $f_D = (1 - Q_{min})(V_{sw}/V_A)$, this range corresponds to effective $f_D$ reductions for these cases from the dispersive bispherical value of $f_D = 0.642$ to $f_D = 0.060$. Table 1 lists these effective values of $f_D$ for each case. The values used to reproduce the observed properties of the outer solar wind lie within this range (Smith et al. 2006).

Our suggestion here that interstellar PUIs are continually heated by the ambient turbulence in the outer solar wind is relatively new, in that previous models of solar wind heating and the large-scale turbulent evolution have not included such an effect (Adhikari et al. 2014; Adhikari et al. 2017; Adhikari et al. 2020; Zank et al. 2018; Oughton et al. 2011; Usmanov & Goldstein 2006; Usmanov et al. 2011; Usmanov et al. 2012, 2014, 2016; Pine et al. 2020d, c, b; Pine et al. 2020a; Ng et al. 2010; Wiengarten et al. 2016; Isenberg 2005; Isenberg et al. 2003, 2010; Smith et al. 2006). However, the observations of PUI out to ~**52** AU from the *New Horizons* mission (Elliot et al. 2019; McComas et al. 2021, 2022) have found that PUIs are not cooled as much as would be predicted by the Vasyliunas & Siscoe (1976) model, which incorporates the effects of adiabatic deceleration under conditions of the radial expansion of a constant-speed solar wind. The *New Horizons* data show that the PUI distribution still generally exhibits a cutoff in speed at the solar wind value in the solar wind frame **(**apart from indications of local shock heating downstream of observed shocks (Zirnstein et al. 2018)), but that the "cooling index" (Chen et al. 2013; Swaczyna et al. 2020; McComas et al. 2021) indicates that the observed PUIs have been heated since their pickup and isotropization. This heating appears to be in addition to, and independent of, the localized energization seen at **these** interplanetary shocks (McComas et al. 2021, 2022).

## 4. A Self-Regulating System

Given the substantial variability of the solar wind, the scenario presented here allows for an intriguing possibility of self-regulation. The ambient turbulence in the expanding wind will decay with increasing heliocentric distance if it is not actively driven by the fluctuations from the PUI instability. If the turbulence in some region is temporarily weak, due perhaps to a faster expansion, the QL scattering can proceed to a bispherical-like shape as originally envisioned by Zank, Matthaeus, and Smith. This interaction will generate fluctuations at the level ~ $V_A/V_{sw}$ times the initial PUI energy, leading to a stronger turbulent driving than that indicated over the large scale by the observations.

In some other region where the turbulence is strong, the turbulent energization will modify the scattering to reduce the QL energy loss from the PUI instability, leading to a corresponding reduction in the turbulent driving. More generally, a parcel of solar wind, advecting outward, could cycle between these two circumstances, actively maintaining a level of turbulent driving which, on average, would be some fraction of the bispherical prediction. A schematic diagram of this regulation is shown in Figure 6.

We do not know the details of the specific mechanism by which the parallel ICWs generated by the PUI instability are transformed into turbulent fluctuations in the distant solar wind. Thus, it is difficult to theoretically establish the fraction of QL intensity that would result from this self-regulation. We show here that a fraction consistent with the observed solar wind heating can be attained within an order-of-magnitude intensity variation of the background turbulence. Future theoretical research and simulation modeling would be extremely valuable in understanding this complicated interaction.

## 5. Summary and Conclusions

Interstellar PUIs, formed from the ionization of inflowing interstellar neutral atoms, dominate the plasma behavior of the supersonic solar wind beyond ~ 40 AU from the Sun. They accumulate with increasing radial position to make up a large fraction of the thermal energy of the outer solar wind, and their properties strongly influence the details of the solar wind termination shock. The energy released when newly ionized PUIs are scattered toward isotropy is thought to drive the observed turbulent heating of the core solar wind protons. However, the interaction predicted from standard QL scattering due to the conventional operation of the anisotropy instability provides far too much energy to the turbulence in the outflowing plasma.

We have suggested that the energizing effects of the ambient background turbulence on the weakly scattering PUIs should be taken into account in the determination of the generated unstable fluctuations. We have illustrated the possible form of this turbulent modification with an idealized computation, using a combined QL diffusive description. We identify a potential self-regulation mechanism in these combined effects, which may explain the observed level of turbulence and solar wind heating in the distant solar wind. Further theoretical work and simulation studies specifically addressing the details of PUI scattering and turbulent driving under realistic conditions are needed to accurately model these interactions in the outer solar wind.

We are grateful for valuable conversations with E. Möbius, P. Swaczyna, and D. Verscharen. PAI, BJV and CWS are supported by NASA grant 80NSSC18K1215. PAI and BJV are also supported by NSF grant AGS2005982. PAI is further supported by NASA grant 80NSSC18K0655.

## References

Adhikari, L., Zank, G. P., Hu, Q., & Dosch, A. 2014, Astrophys. J., 793, 52
Adhikari, L., Zank, G. P., Zhao, L.-L., & Webb, G. M. 2020, Astrophys. J., 891, 34
Adhikari, L., Zank, G. P., Hunana, P., et al. 2017, Astrophys. J., 841, 85
Aggarwal, P., Taylor, D. K., Smith, C. W., et al. 2016, Astrophys. J., 822, 94
Argall, M. R., Fisher, M. K., Joyce, C. J., et al. 2015, Geophys. Res. Lett., 42, 9617
Argall, M. R., Hollick, S. J., Pine, Z. B., et al. 2017, Astrophys. J., 849, 61
Axford, W. I. 1972, in Solar Wind, ed. C. P. Sonett, P. J. Coleman, & J. M. Wilcox (Washington, D.C.: NASA Spec. Pub. SP-308), 609
Brodiano, M., Andrés, N., & Dmitruk, P. 2021, Astrophys. J., 922, 240
Cannon, B. E., Smith, C. W., Isenberg, P. A., et al. 2014a, Astrophys. J., 784, 150
Cannon, B. E., Smith, C. W., Isenberg, P. A., et al. 2014b, Astrophys. J., 787, 133
Chen, J. H., Möbius, E., Gloeckler, G., et al. 2013, J. Geophys. Res., 118, 3946
Cowee, M. M., Russell, C. T., & Strangeway, R. J. 2008, J. Geophys. Res., 113, A08220
Elliott, H. A., McComas, D. J., Zirnstein, E. J., et al. 2019, Astrophys. J., 885, 156
Fisher, M. K., Argall, M. R., Joyce, C. J., et al. 2016, Astrophys. J., 830, 47

Florinski, V., Heerikhuisen, J., Niemiec, J., & Ernst, A. 2016, Astrophys. J., 826, 197
Florinski, V., Zank, G. P., Heerikhuisen, J., Hu, Q., & Khazanov, I. 2010, Astrophys. J., 719, 1097
Galeev, A. A., & Sagdeev, R. Z. 1988, Astrophys. Space Sci., 144, 427
Gary, S. P. 1993, Theory of Space Plasma Microinstabilities (Cambridge, England:
Gazis, P. R. 1984, J. Geophys. Res., 89, 775
Gazis, P. R., & Lazarus, A. J. 1982, Geophys. Res. Lett., 9, 431
Goldreich, P., & Sridhar, S. 1995, Astrophys. J., 438, 763
———. 1997, Astrophys. J., 485, 680
Hellinger, P., & Trávnícek, P. 2015, J. Plasma Phys., 81, 305810103
———. 2016, Astrophys. J., 832, 32
Hellinger, P., Matteini, L., Landi, S., et al. 2015, Astrophys. J., 811, L32
Hollick, S. J., Smith, C. W., Pine, Z. B., et al. 2018a, Astrophys. J., 863, 75
———. 2018b, Astrophys. J., 863, 76
Holzer, T. E. 1972, J. Geophys. Res., 77, 5407
Huddleston, D. E., Coates, A. J., & Johnstone, A. D. 1992, J. Geophys. Res., 97, 19,163
Huddleston, D. E., Strangeway, R. J., Warnecke, J., Russell, C. T., & Kivelson, M. G. 1998, J. Geophys. Res., 103, 19887
Isenberg, P. A. 1986, J. Geophys. Res., 91, 9965
Isenberg, P. A. 2005, Astrophys. J., 623, 502
Isenberg, P. A., & Lee, M. A. 1996, J. Geophys. Res., 101, 11,055
Isenberg, P. A., & Vasquez, B. J. 2019, Astrophys. J., in press, arXiv: 1910.11366
Isenberg, P. A., Smith, C. W., & Matthaeus, W. H. 2003, Astrophys. J., 592, 564
Isenberg, P. A., Smith, C. W., Matthaeus, W. H., & Richardson, J. D. 2010, Astrophys. J., 719, 716
Johnstone, A. D., Huddleston, D. E., & Coates, A. J. 1991, in Cometary Plasma Processes, ed. A. D. Johnstone (Washington, DC: AGU), 259
Joyce, C. J., Smith, C. W., Isenberg, P. A., Murphy, N., & Schwadron, N. A. 2010, Astrophys. J., 724, 1256
Kolmogorov, A. N. 1941a, C. R. Acad. Sci. URSS, 32, 16
———. 1941b, C. R. Acad. Sci. URSS, 30, 301
Lee, M. A. 1971, Plasma Phys., 13, 1079
Lee, M. A., & Ip, W.-H. 1987, J. Geophys. Res., 92, 11,041
Liu, K., Möbius, E., Gary, S. P., & Winske, D. 2012, J. Geophys. Res., 117, A10102
Marchuk, A. V., Smith, C. W., Watson, A. S., et al. 2021, Astrophys. J., 923, 185
Markovskii, S. A., & Vasquez, B. J. 2022, Astrophys. J., 924, 111
Markovskii, S. A., Vasquez, B. J., & Chandran, B. D. G. 2019, Astrophys. J., 875, 125
———. 2020, Astrophys. J., 889, 7
Matthaeus, W. H., Zank, G. P., Smith, C. W., & Oughton, S. 1999, Phys. Rev. Lett., 82, 3444
McComas, D.J., Shrestha, B. L., Swaczyna, P., et al. 2022, Astrophys. J., 934, 147
McComas, D. J., Swaczyna, P., Szalay, J. R., et al. 2021, Astrophys. J. Suppl., 254, 19
McComas, D. J., Zirnstein, E. J., Bzowski, M., et al. 2017, Astrophys. J. Suppl., 233, 8
McComas, D. J., Allegrini, F., Bagenal, F., et al. 2008, Space Sci. Rev., 140, 261
Min, K., & Liu, K. 2018, Astrophys. J., 852, 39
Min, K., Liu, K., & Gary, S. P. 2017, J. Geophys. Res., 122, 7891

Murphy, N., Smith, E. J., Tsurutani, B. T., Balogh, A., & Southwood, D. J. 1995, Space Sci. Rev., 72, 447
Ng, C. S., Bhattacharjee, A., Munsi, D., Isenberg, P. A., & Smith, C. W. 2010, J. Geophys. Res., 115, A02101
Oughton, S., Matthaeus, W. H., Wan, M., & Osman, K. T. 2015, Phil. Trans. R. Soc. A, 373, 201401152
Oughton, S., Matthaeus, W. H., Smith, C. W., Breech, B., & Isenberg, P. A. 2011, J. Geophys. Res., 116, A08105
Pine, Z. B., Smith, C. W., Hollick, S. J., et al. 2020a, Astrophys. J., 900, 94
Pine, Z. B., Smith, C. W., Hollick, S. J., et al. 2020b, Astrophys. J., 900, 93
———. 2020c, Astrophys. J., 900, 92
———. 2020d, Astrophys. J., 900, 91
Schlickeiser, R. 1989, Astrophys. J., 336, 243
Smith, C. W., Isenberg, P. A., Matthaeus, W. H., & Richardson, J. D. 2006, Astrophys. J., 638, 508
Smith, C. W., Matthaeus, W. H., Zank, G. P., et al. 2001, J. Geophys. Res., 106, 8253
Swaczyna, P., McComas, D. J., Zirnstein, E. J. et al. 2020, Astrophys. J., 903, 48
Summerlin, E. J., Viñas, A. F., Moore, T. E., Christian, E. R., & Cooper, J. F. 2014, Astrophys. J., 793, 93
Szegö, K., Glassmeier, K.-H., Bingham, R., et al. 2000, Space Sci. Rev., 94, 429
TenBarge, J. M., & Howes, G. G. 2012, Phys. Plasmas, 19, 055901
Usmanov, A. V., & Goldstein, M. L. 2006, J. Geophys. Res., 111, A07101
Usmanov, A. V., Goldstein, M. L., & Matthaeus, W. H. 2012, Astrophys. J., 754, 40
———. 2014, Astrophys. J., 788, 43
———. 2016, Astrophys. J., 820, 17
Usmanov, A. V., Matthaeus, W. H., Breech, B., & Goldstein, M. L. 2011, Astrophys. J., 727, 84
Vasquez, B. J., Smith, C. W., Hamilton, K., MacBride, B. T., & Leamon, R. J. 2007, J. Geophys. Res., 112, A07101
Vasyliunas, V. M., & Siscoe, G. L. 1976, J. Geophys. Res., 81, 1247
von Kármán, T., & Howarth, L. 1938, Proc. R. Soc. London, Ser. A, 164, 192
Wiengarten, T., Oughton, S., Engelbrecht, N. E., et al. 2016, Astrophys. J., 833, 17
Williams, L. L., & Zank, G. P. 1994, J. Geophys. Res., 99, 19,229
Williams, L. L., Zank, G. P., & Matthaeus, W. H. 1995, J. Geophys. Res., 100, 17,059
Wu, C. S., & Davidson, R. C. 1972, J. Geophys. Res., 77, 5399
Yang, L., Zhang, L., He, J., et al. 2018, Astrophys. J., 866, 41
Zank, G. P., Matthaeus, W. H., & Smith, C. W. 1996, J. Geophys. Res., 101, 17,093
Zank, G. P., Adhikari, L., Zhao, L.-L., et al. 2018, Astrophys. J., 869, 23
Zirnstein, E. J., McComas, D. J., Kumar, R., et al. 2018, PhRvL, 121, 075102

**Table 1.** Minimum Values of the Normalized Energy Loss, $Q$, and the Corresponding Effective Values of $f_D$ for the Cases of Section 3.

| $A_{turb}$ | $Q_{min}$ | eff $f_D$ |
|---|---|---|
| 0.0 | 0.893 | 0.642 |
| 0.0001 | 0.910 | 0.540 |
| 0.0005 | 0.945 | 0.330 |
| 0.001 | 0.962 | 0.228 |
| 0.002 | 0.984 | 0.096 |
| 0.003 | 0.990 | 0.060 |

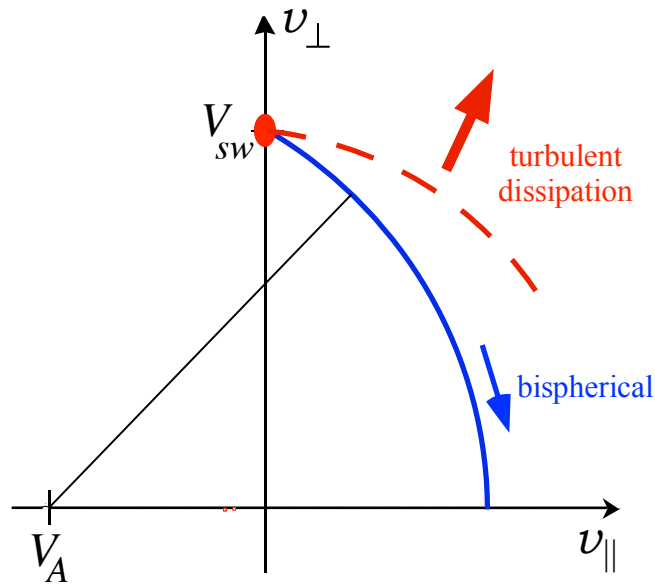

**Figure 1.** Sketch of the plasma response of newly ionized PUIs in the azimuthal magnetic field of the outer solar wind. The ions initially form an unstable ring distribution, depicted by the red dot, which scatters to a closed shell in phase space. The conventional QL theory predicts a bispherical shell, given by the blue curve as reflected into the other quadrants of the figure. The red dashed curve indicates a possible modification of the shell due to interaction with the ambient turbulence.

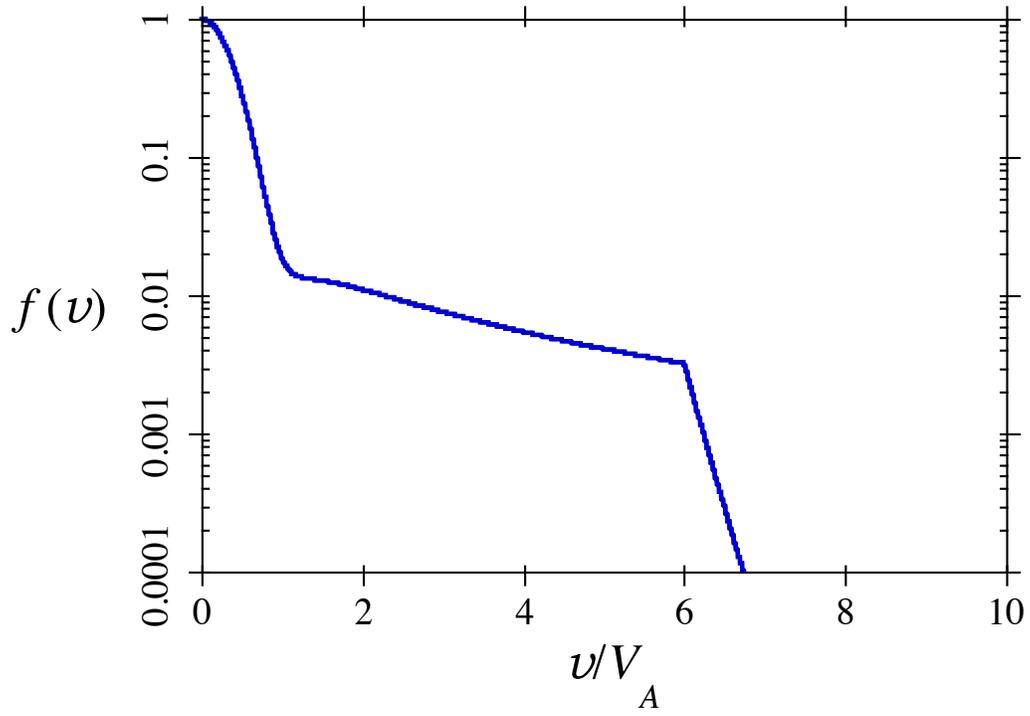

**Figure 2.** Normalized initial proton distribution function (8).

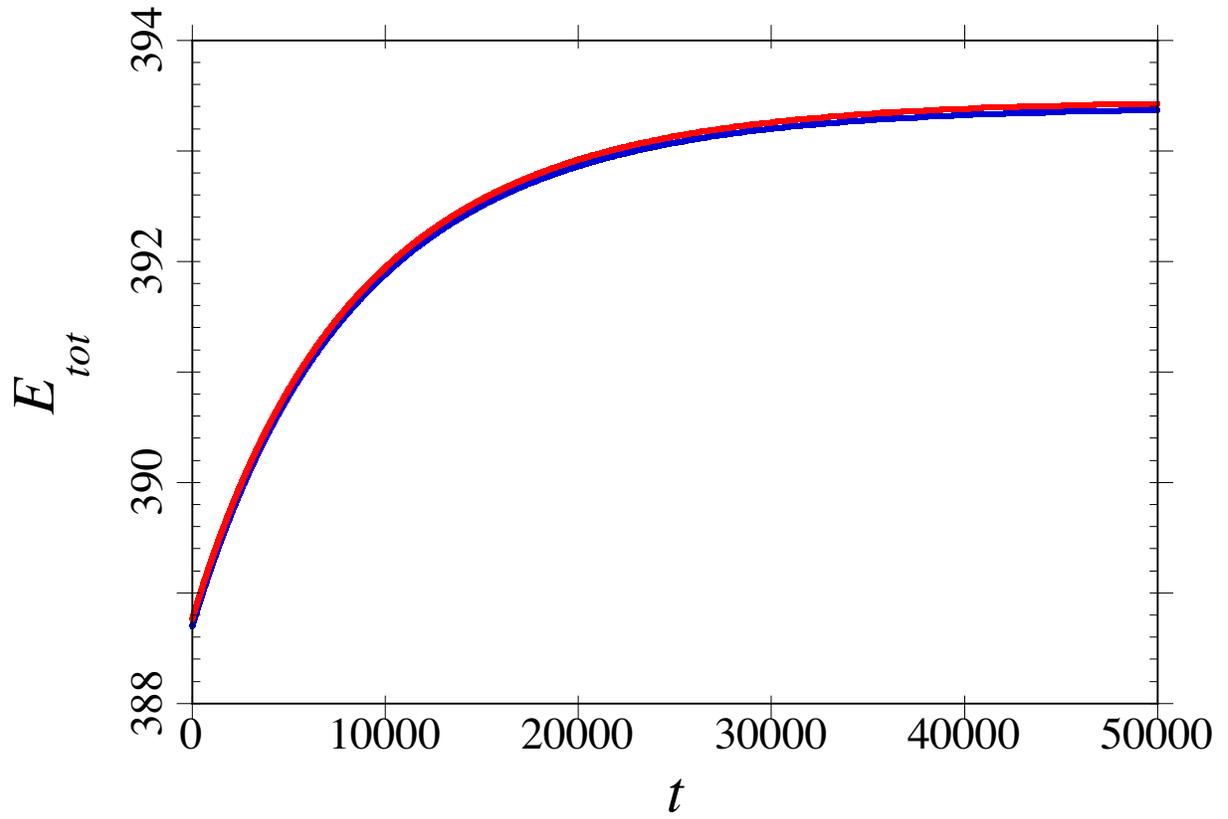

**Figure 3.** Evolution of the total proton energy in grid units, when $A_{turb} = 0$. The blue curve uses the initial distribution (8), and the red curve has the additional slug of new PUIs.

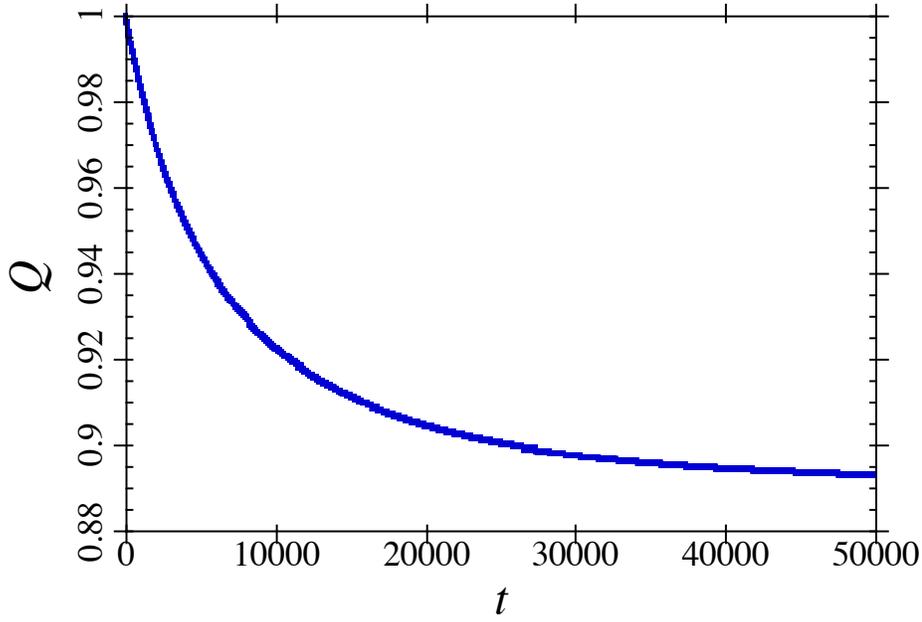

**Figure 4.** Evolution of the normalized energy difference, $Q$, between the computations with and without the slug of new PUIs, for $A_{turb} = 0$.

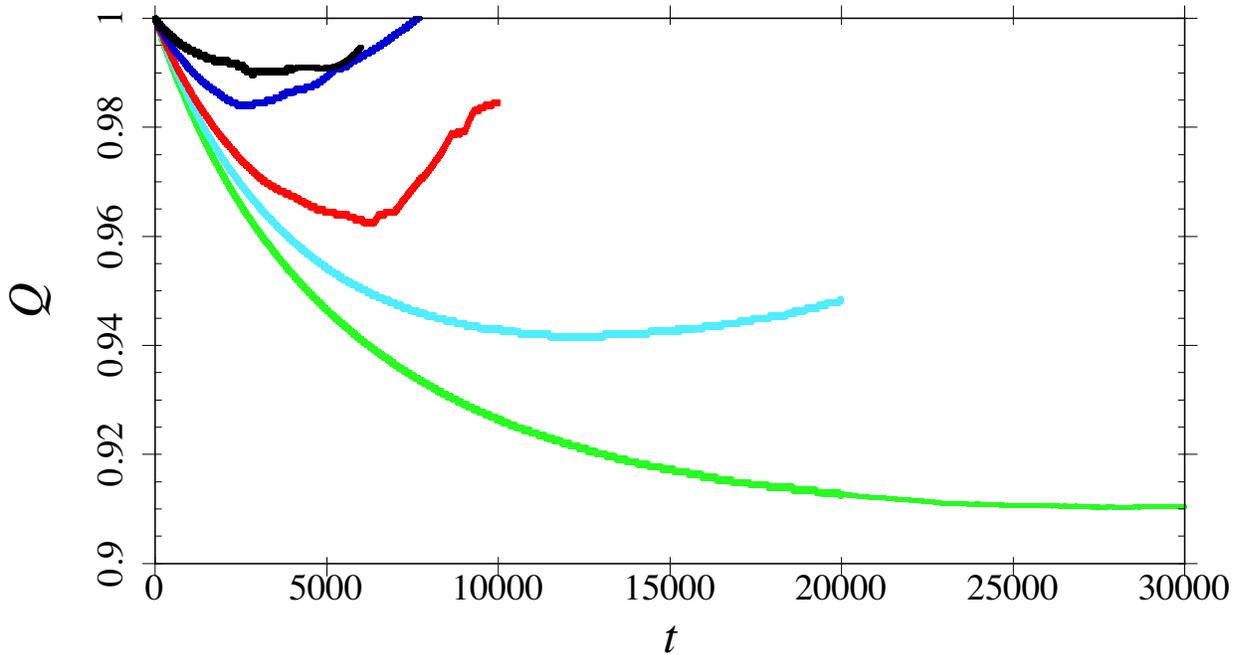

**Figure 5.** Evolution of the normalized energy difference, $Q$, for increasing values of $A_{turb}$, while $A_{IC} I(k_o)$ is fixed at $10^{-5}$. The values are $A_{turb} = 0.0001$ (green), $0.0005$ (light blue), $0.001$ (red), $0.002$ (dark blue), and $0.003$ (black). The respective minima are given in Table 1.

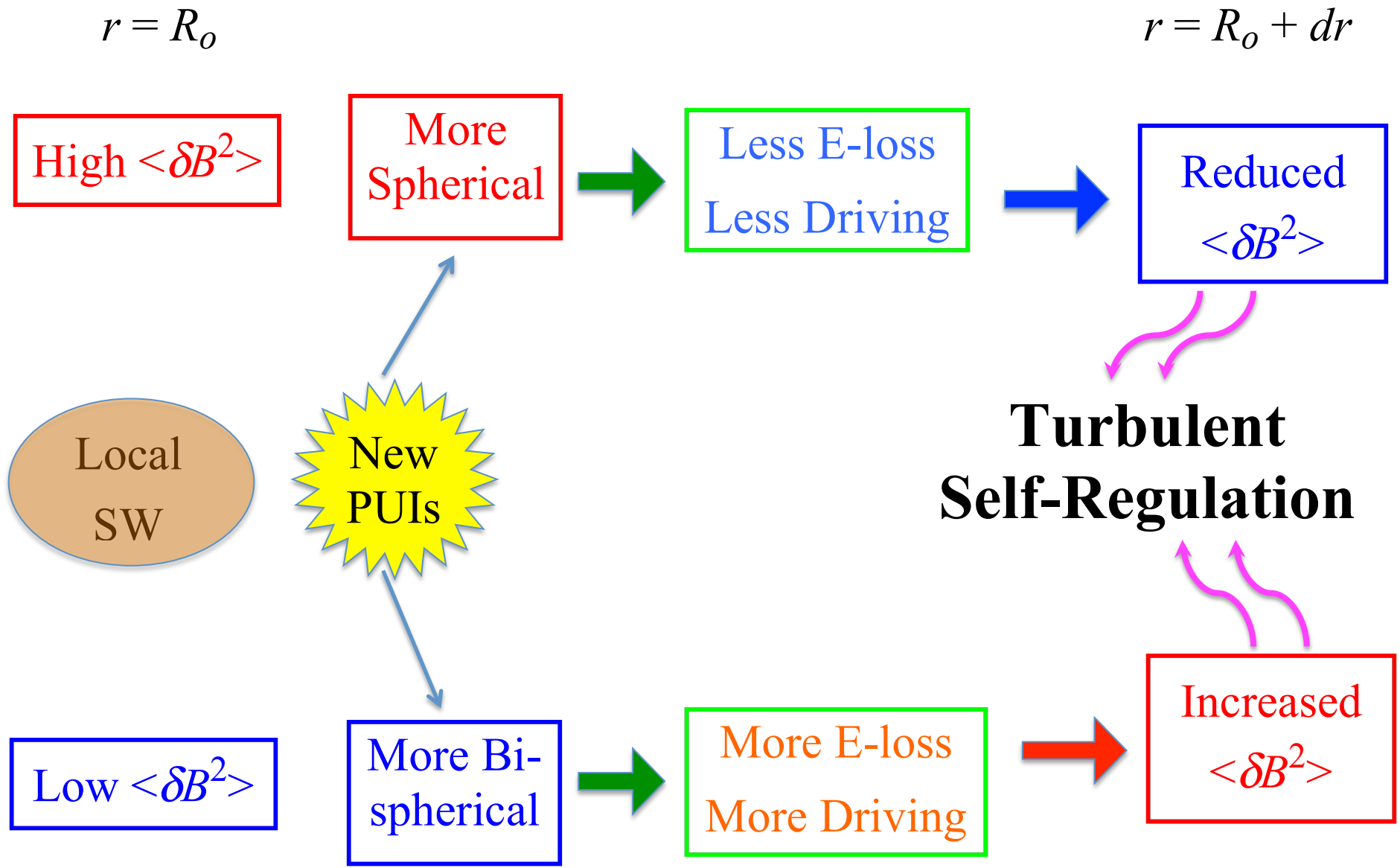

**Figure 6.** Schematic diagram of the self-regulation mechanism.